\documentclass[english, aps, prb, twocolumn, superscriptaddress, secnumarabic, showkeys, nofootinbib, floatfix]{revtex4-1}

\usepackage[english]{babel}

\usepackage{amssymb,amsmath}
\usepackage{enumerate}
\usepackage{grffile}
\usepackage{balance,float}

\usepackage[pdftex]{graphicx}
\usepackage[caption=false]{subfig}
\usepackage{color}
\usepackage[usenames,dvipsnames]{xcolor}
\usepackage[colorlinks=true,linkcolor=RoyalBlue,citecolor=PineGreen,urlcolor=MidnightBlue,linktoc=page]{hyperref}

\newcommand{\ps}[1]{#1_\mathrm{ps}}

\newcommand{\jc}{j_\mathrm{c}}
\newcommand{\jn}{j_\mathrm{n}}
\newcommand{\js}{j_\mathrm{s}}
\newcommand{\jdp}{j_\mathrm{dp}}
\newcommand{\jr}{j_\mathrm{r}}
\newcommand{\jext}{j}
\newcommand{\Tc}{T_\mathrm{c}}

\newcommand{\D}{\mathcal{D}}
\renewcommand{\u}{\mathfrak{u}}

\renewcommand{\Im}{\mathrm{Im}}

\newcommand{\IV}{$I$--$V \,$}

\begin{document}

\title{Instabilities of the normal state in current-biased narrow superconducting strips}

\author{Yury N. Ovchinnikov}
\affiliation{Landau Institute for Theoretical Physics, RAS, Chernogolovka, Moscow District, 142432, Russia}
\affiliation{Max-Plank Institute for Physics of Complex Systems, 01187 Dresden, Germany}

\author{Andrey A. Varlamov}
\affiliation{CNR-SPIN (Istituto Superconduttori, Materiali Innovativi e Dispositivi), Viale del Politecnico 1, 00133 Rome, Italy} 

\author{Gregory J. Kimmel}
\affiliation{Materials Science Division, Argonne National Laboratory, 9700 S Cass Avenue, Lemont, IL 60439, USA}

\author{Andreas Glatz}
\affiliation{Materials Science Division, Argonne National Laboratory, 9700 S Cass Avenue, Lemont, IL 60439, USA}
\affiliation{Department of Physics, Northern Illinois University, DeKalb, IL 60115, USA}

\date{\today}

\begin{abstract}
We study the current-voltage characteristic of narrow superconducting strips in the gapless regime near the critical temperature in the framework of the Ginzburg-Landau model.
Our focus is on its instabilities occurring at high current biases. 
The latter are consequences of dynamical states with periodic phase-slip events in space and time. 
We analyze their structure and derive the value of the reentrance current at the onset of the instability of the normal state.
It is expressed in terms of the kinetic coefficient of the time-dependent Ginzburg-Landau equation and calculated numerically. 
\end{abstract}


\maketitle

\section{Introduction}

Narrow superconducting strips are the subject of great interest for superconducting quantum electronic devices. Their dissipation-less state is very subtle and sensitive to thermal and quantum fluctuations, which can easily flip the superconducting strip into the resistive state, making them ideal candidates for very sensitive detectors. 
Various models have been proposed to explain the appearance of non-zero resistance in these strips and its temperature dependence in the region of low temperatures (for a review, see Refs.~[\onlinecite{Sahu:2009,Bezryadin:2012}]).

The role of thermal fluctuations responsible for energy dissipation, when current flows through a one-dimensional superconductor, was considered for the first time in the seminal paper by Langer and Ambegaokar\cite{Langer:1967} over fifty years ago. Note, that a realistic ``one-dimensional superconductor'' is in fact a narrow strip with finite width $W$, much less than the Ginzburg-Landau coherence length $\xi(\tau)\propto (k_{\mathrm{B}}\Tc\tau)^{-1/2}$, where $\tau=1-T/\Tc$ is the reduced temperature and $\Tc$  the critical temperature. 
The energy dissipation in this system is related to phase-slip processes appearing in thin superconducting wires~\cite{tinkham1996introduction,Tinkham1974PSC,Quasi-1D-Superconductors,kramer1984structure,rangel1989theory,lau2001quantum,mooij2006superconducting,mckay2008phase,Kimmel:2017b} or superfluids~\cite{glatz+prl02,Scherpelz:2014,scherpelz+pra15}, i.e., the processes of vortices/flux quanta crossing the strip. 

It is clear, that such events cannot be realized in the framework of a purely one-dimensional model. Indeed, as it was shown in Ref.~[\onlinecite{Langer:1967}], the minimal value of the order parameter magnitude remains finite and equal to $\left( 2/3\right) ^{1/2}\Delta_{ \mathrm{BCS}}$ even when the density of current  flowing through the one-dimensional superconductor reaches the ``depairing'' value, $\jdp\propto e\nu\left(  k_{\mathrm{B}}T\tau\right)^{2}\xi(\tau)$, and 
global  superconductivity in the one-dimensional channel  is only partially suppressed.
Yet,  the order parameter should become zero at least in one point of the strip in order to allow the system to perform a phase slip event. \footnote{Note, that, in contrast to the one-dimensional superconducting strip, the order parameter in a higher-dimensional superconductor becomes zero when the current density approaches the  ``depairing'' value. }

The apparent paradox occurring in the one-dimensional case was resolved in Ref.~[\onlinecite{Ovchinnikov:2015}]. 
The authors demonstrated  that the saddle point solutions of the static Ginzburg-Landau (GL) equations for the order parameter $\widetilde{\psi}$ in the presence of a fixed current density $\jext$, possessing at least one vortex, exist only for very weak current densities  $\jext<\jc=0.0312 (W/\xi)\jdp$. 
In the case  $\jext>\jc$, such phase-slip events are possible, which are random in space and time due to thermal fluctuations.
When the current density exceeds the value $\jdp$,  the static scenario described above does not hold anymore.   
In this region ($\jext>\jdp$)  dynamical states are formed in the strip, where phase-slip events occur periodically in space and time \footnote{Note that material defects and weak links also lead to phase-slip events~\cite{Kramer:1978,Kramer:1981,Berdiyorov:2014,Kimmel:2017b}, but here we concentrate on the clean case.}.

In this paper we  study the hysteretic structure of such dynamical states of a narrow superconducting strip and obtain the corresponding current-voltage  (\IV) characteristics.  
In particular, we  derive the value of the re-entrance current density $\jr$, at the onset of the instability of the normal state, when the applied current decreases.
Our consideration is valid in the gapless region, at temperatures slightly below the critical one,  where the time-dependent Ginzburg-Landau equation holds. 
In this situation, we investigate the strong current regime $\jext>\jdp$ of the superconducting strip being in its dynamical state up to second order perturbation theory in the electric field $E$. 

It is important to note, that the considered situation is quite different from the problem of determining the critical current at which the superconductor becomes normal (see Refs.~[\onlinecite{Ivlev:1984,Kopnin:1984}]). 
The authors of the cited papers, which are based on the works by Gorkov~\cite{Gorkov:1970} and Kulik~\cite{Kulik:1971}, claim that the normal state remains stable for any finite value of $E$. 
However, they ignore the fact that the exponential growth of superconducting fluctuations in a time interval determined by $\jext \Delta t <\hbar \sigma /[e\xi(\tau)]$ leads to an instability of the normal state ($\sigma$ is the normal conductivity).  
As a result, the system enters a dynamical superconducting state at some finite electric field. We will show below that there are many values of the parameter $\u E$ for which the normal state starts to be unstable even for infinitely small perturbations.

In the following we describe the model and show the analysis of the time-dependent Ginzburg-Landau equation near the critical point of the normal state.
We derive the value of the reentrance current and order parameter values using first order perturbation theory by small deviation from the critical point. Details of the calculations can be found in the Appendices.
We start with introducing the model in the following section and then analyze the \IV characteristics close to the deparing current and near the instability points of the normal state.

\section{Model} \label{sec:GL}

\begin{figure}
\includegraphics[width=0.9\columnwidth]{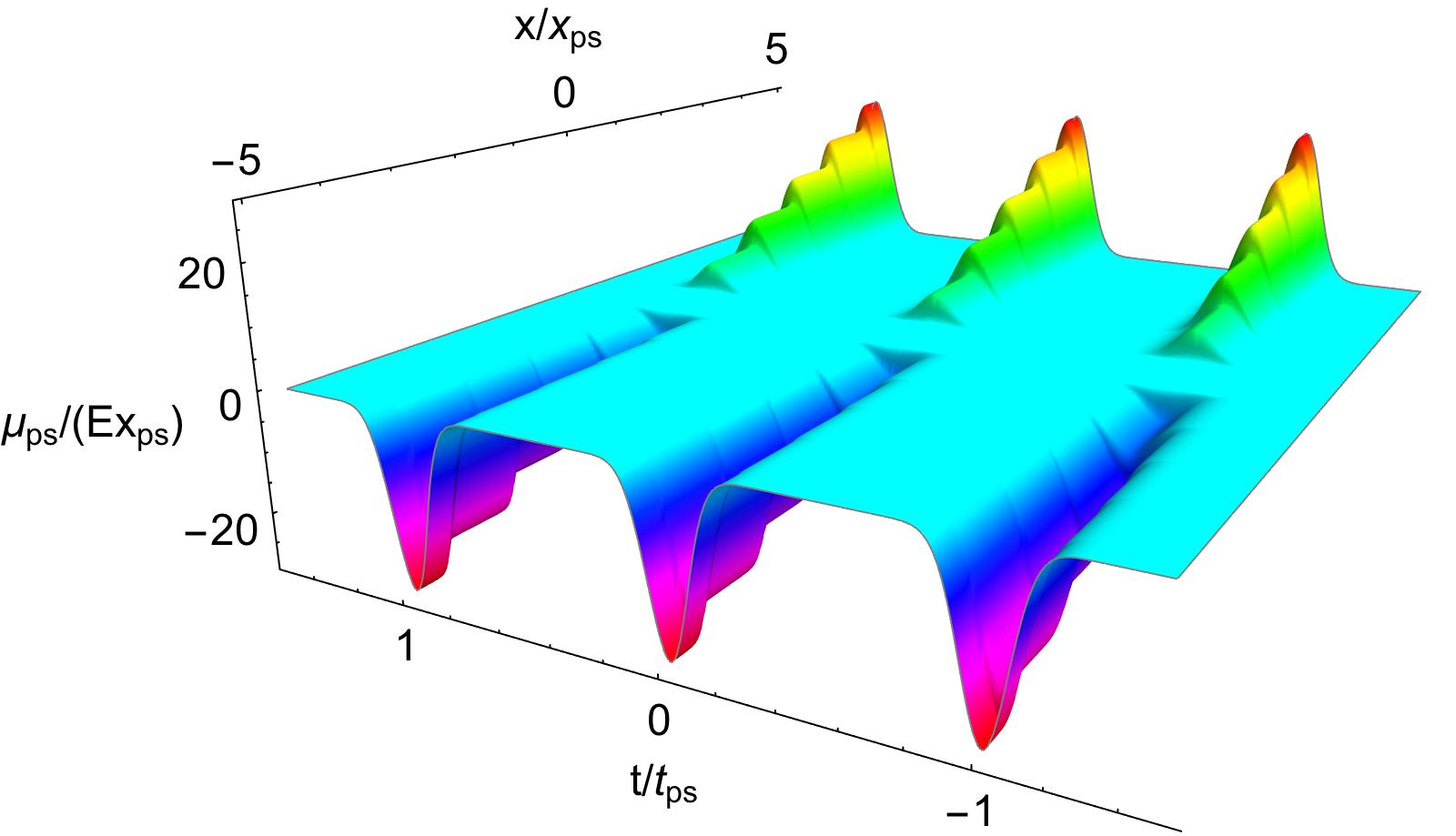}\\
\includegraphics[width=0.45\columnwidth]{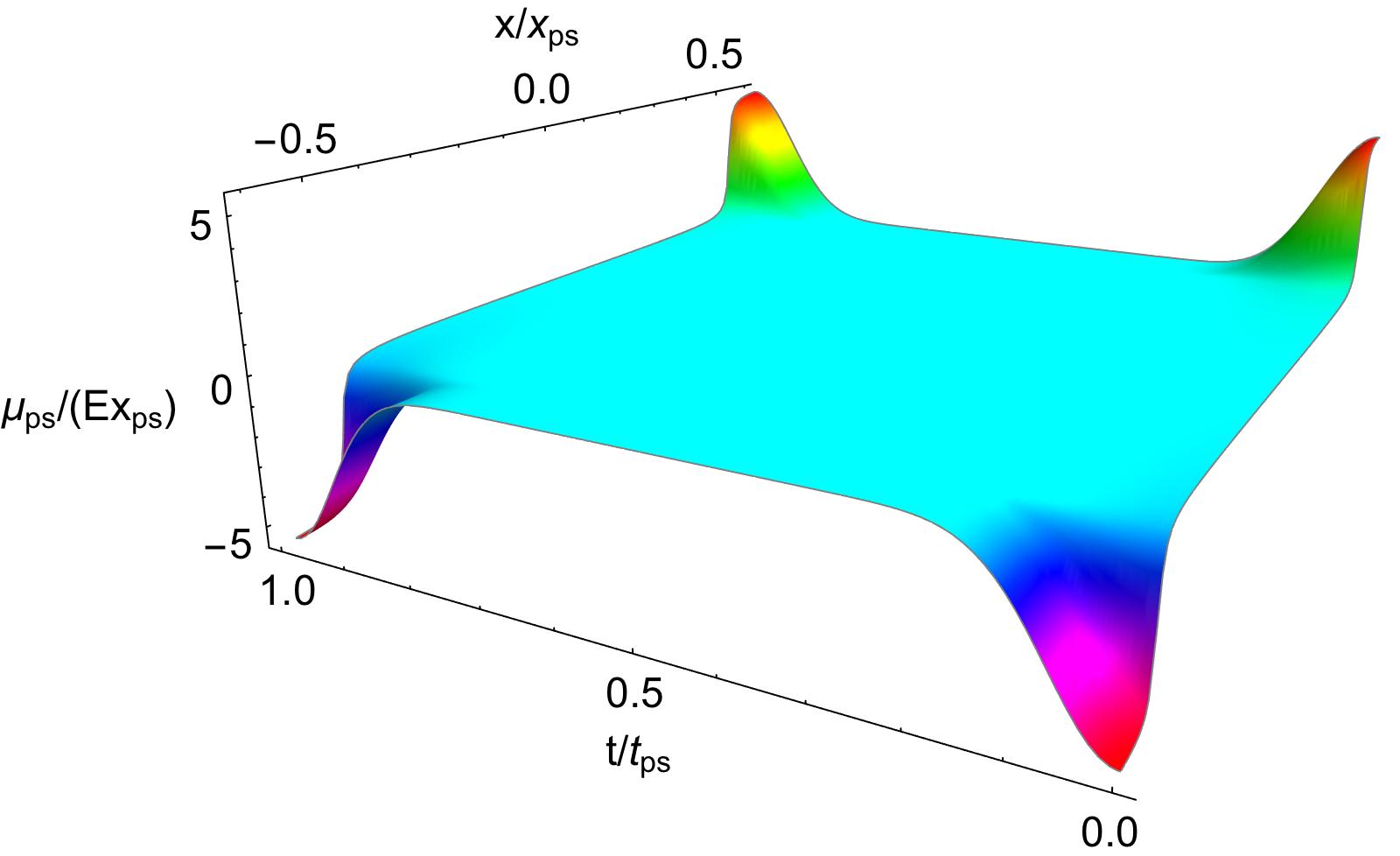}
\includegraphics[width=0.45\columnwidth]{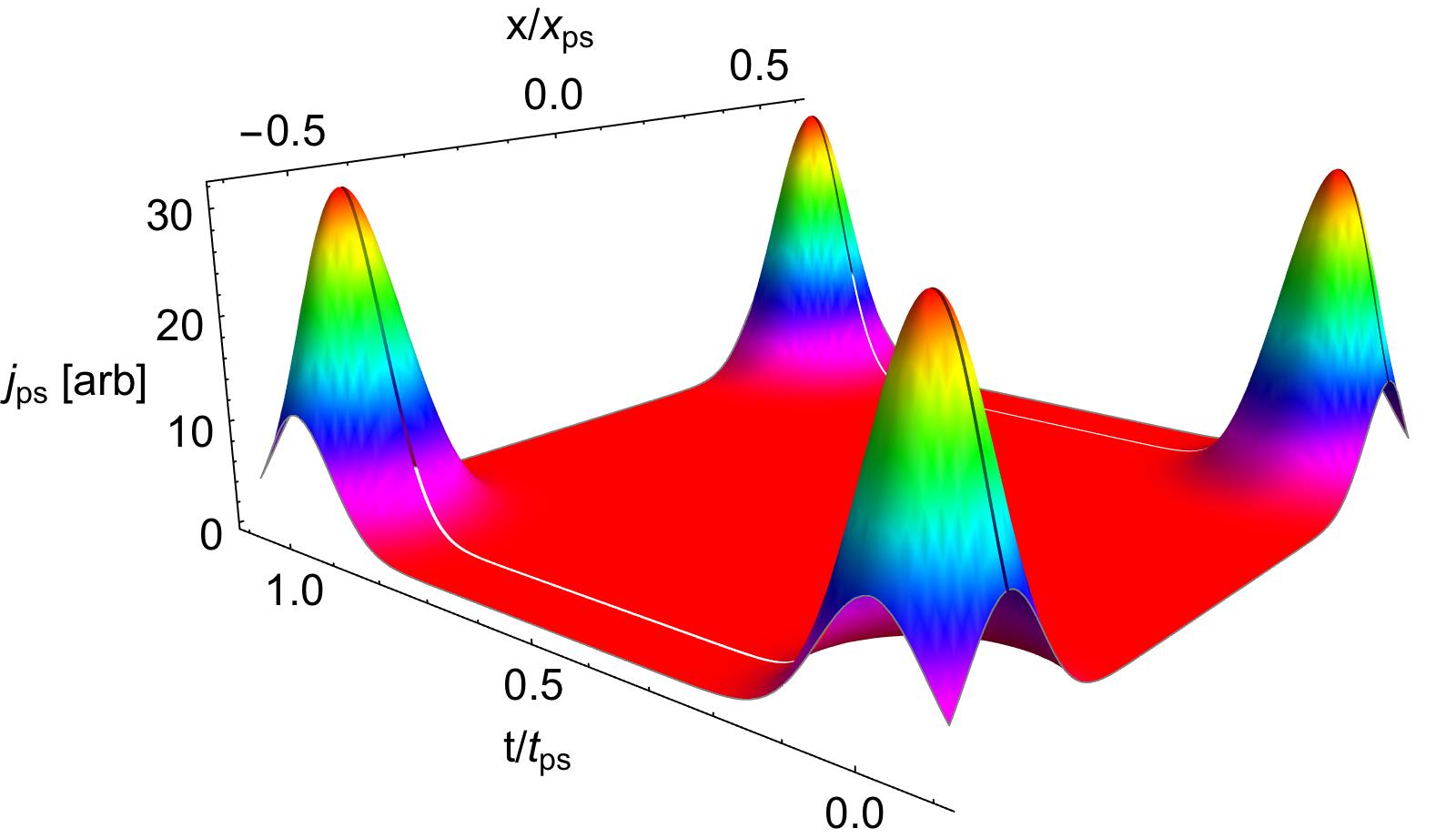}
\caption{Illustration of phase-slip potential $\ps{\mu}$, Eq.~\eqref{eq:mutilde}, as function of time and space and finite width delta functions (top), see text. \textit{Bottom:} close-up of region $\ps{\mathcal{A}}$ (left) and related $\ps{j}$ (right) in the same region.}
\label{fig:mups}
\end{figure}

In this paper we approximate the narrow superconducting strip of width $W$ smaller than the superconducting coherence length by a one-dimensional (1D) system, described by the time-dependent Ginzburg-Landau equation (TDGLE). 
The TDGLE can be written in dimensionless variables, without accounting for thermal fluctuations and magnetic field (the latter does not appear in the 1D model) in the form
\begin{equation}
\u\left(\partial_t+i\mu\right)\psi=\partial_x^2\psi+\left(1-|\psi|^2\right)\psi,
\label{eq:TDGLE}
\end{equation}
where $\psi$ is the complex order parameter and $\mu$ is the scalar potential.  
The reduced relaxation rate $\u$ controls the system's evolution in time and is given by 
\begin{equation}
\label{eq:mu}
\u= \u_0\frac{\pi^4}{14\zeta(3)}\frac{e^2\nu\mathfrak{D}}{\sigma}\,,
\end{equation}
where $\nu$ is the density of states at the Fermi surface, $\mathfrak{D}$ is the effective diffusion constant, $\zeta$ is the Riemann-Zeta function, and $\u_0$ is a numerical constant.\cite{Ovchinnikov1997,KopninBook_2001}

The fixed total bias-current consists of the sum of normal ($\jn$) and superconducting ($\js$) components and in its turn can be related to the space derivatives of the complex order parameter and  the scalar potential:
\begin{equation}
\jext=\jn+\js=-\partial_x\mu-\imath\Im(\psi^*\partial_x\psi)\,.
\label{eq:current}
\end{equation}
Time and distance in Eqs. (\ref{eq:TDGLE})-(\ref{eq:current})  are measured in units of $t_{0} = 8\pi \sigma \lambda^{2} / c^{2}$ and scaled superconducting coherence length, $\xi=\left(\frac{\pi\mathfrak{D}}{16k_B\Tc\tau}\right)^{1/2}$, respectively, with $\lambda$ being the London penetration depth. 
The electrical current density $j$ is measured in units of $j_{0} = c\Phi_{0}/(8\pi^{2}\lambda^{2}\xi \sqrt 2)$ ($\Phi_0$ is the flux quantum).
In these units the depairing current density reads as  $\jdp = 2/(3\sqrt{3})j_{0} \approx 0.385 j_{0}$.

In what follows, it is useful to transform the scalar potential to the form
\begin{equation}
\label{eq:mu}
\mu=-Ex+\tilde{\mu}\,,
\end{equation}
where $E$ is the average electric field, which is equal to the average normal current  in dimensionless units, and $\tilde{\mu}$ is the spatially and temporally fluctuating part of $\mu$.

Phase-slip events generate instabilities in the current-voltage characteristics, which will be the main subject of this work.
In particular, these processes cause strong changes in the electric field and order parameter at their space-time coordinates.
We take their effect into account explicitly by introducing a corresponding effective electric potential $\tilde{\mu}$ to the TDGLE. 

A phase-slip process in a strip of finite width is related to the transfer of a magnetic vortex-antivortex pair across it. 
Each such event is accompanied by the suppression of the order parameter and, consequently, of the supercurrent. 
Due to the conservation of the total current, a sharp peak in the normal current appears at the time and space location of the phase-slip event.

In the large-current regime, $\jext>\jdp$, these phase-slip events (ps) are periodic in time and space.  
The corresponding periods in time and space are denoted as $\ps{t}$ and $\ps{x}$, respectively. 
Outside a very narrow region in the time-space plane one can rewrite Eq.~\eqref{eq:TDGLE} in the form
\begin{equation}
\u\left[\partial_t+\imath(\tilde{\mu}+\ps{\mu}-E\tilde{x})\right]\psi=\partial_x^2\psi+\left(1-|\psi|^2\right)\psi.
\label{eq:TDGLEPS}
\end{equation}
Here  $\tilde{x}=x\mod (\ps{x}/2)$ and the associated potential $\ps{\mu}$
\begin{eqnarray}
&&\ps{\mu}= E \ps{x} \ps{t}\sum_k\delta(t-\ps{t}k)\times\nonumber\\
&&\sum_m m\Theta\left[x-\ps{x}\left(m-\frac{1}{2}\right)\right]\Theta\left[\ps{x}\left(m+\frac{1}{2}\right)-x\right]\!, \label{eq:mutilde}
\end{eqnarray}
where the quantization condition  $E\ps{x}\ps{t}=2\pi$ is implied.
The shape of $\ps{\mu}$ is illustrated in Fig.~\ref{fig:mups} (top) with smoothed step and delta functions. 
In the bottom panels $\ps{\mu}$ and $\ps{j}\propto \partial_x\ps{\mu}$ are shown in the elementary space-time cell $\ps{\mathcal{A}}\equiv ]-\ps{x}/2;\ps{x}/2[\times ]0;\ps{t}[$. The latter illustrating the mentioned spikes in the normal current at the phase slip event location.

These additional terms do not contribute to the electric field inside $\ps{\mathcal{A}}$ but only at its corners. 
The solution for the order parameter $\psi$ away from the corners of $\ps{\mathcal{A}}$ follows from Eq.~\eqref{eq:TDGLEPS} with periodic boundary conditions. 
The size of the region with strong suppression of $\psi$ is of order $W\times W\ps{t}/\ps{x}$, where $W$ is the width of the strip.

\section{Analysis of the \IV characteristics}
Let us start with the analysis of Eqs.~\eqref{eq:TDGLE}-\eqref{eq:mu}. 
Within the unit cell $\ps{\mathcal{A}}$, one can make the Fourier-Ansatz for the complex order parameter 
\begin{equation}
\psi=\sum_{k=-\infty}^\infty A_k e^{\imath Q k x},
\label{eq:psiansatz}
\end{equation}
where $Q$ should be found from local minimum conditions of $E$ for a given current density $\jext$.

Correspondingly, Eq.~\eqref{eq:current} acquires the form
\begin{equation}
	j=E+Q\sum_{k}k|A_k|^2\,.
	\label{eq:jansatz}
\end{equation}

Plugging Ansatz~\eqref{eq:psiansatz} into Eq.~\eqref{eq:TDGLE} gives
\begin{widetext}
\begin{eqnarray}
 &  & -\imath\u\left\{ Ex+\frac{\imath}{2}\sum_{k\neq0}k^{-1}[A_{1}(2-k)A_{1-k}^{*}+A_{1}^{*}(2+k)A_{k+1}]e^{\imath kQx}-\frac{\imath}{2}\sum_{l\neq k\neq1}\frac{k+l}{k-l}A_{k}^{*}A_{l}e^{-\imath\left(k-l\right)Qx})\right\} \sum_{k=-\infty}^{\infty}A_{k}e^{\imath kQx} \nonumber \\
\nonumber \\
 &  & =(1-|A_{1}|^{2}-Q^{2})A_{1}e^{\imath Qx}+\sum_{k\neq 1}\left\{ (1-2|A_{1}|^{2}-k^{2}Q^{2})A_{k}e^{\imath kQx}-A_{1}^{2}A_{k}^{*}e^{-\imath(k-2)Qx}\right.\nonumber \\
 &  & \left.-\sum_{l\neq1}\left[2A_{1}A_{k}A_{l}^{*}e^{\imath(k-l+1)Qx}+A_{1}^{*}A_{k}A_{l}e^{\imath(k+l-1)Qx}+\sum_{m\neq1}A_{k}A_{l}A_{m}^{*}e^{\imath(k+l-m)Qx}\right]\right\}. \nonumber \\
\label{eq:FTTDGLE}
\end{eqnarray}
Finally, accounting for the charge conservation condition $\textrm{div} \jext=0$  and Eq.~\eqref{eq:current}, one can find the expression for the fluctuating part of the scalar potential  $\tilde{\mu}$ in terms of the introduced Fourier coefficients:
\begin{equation}
\tilde{\mu} =  -\frac{\imath}{2}\left\{ \sum_{k\neq0}\frac{e^{\imath kQx}}{k}\left[A_{1}(2-k)A_{1-k}^{*}+A_{1}^{*}A_{k+1}(2+k)\right]-\sum_{k\neq l\neq1}\left(\frac{k+l}{k-l}\right)e^{-\imath(k-l)Qx}A_{k}^{*}A_{l}\right\}\,. \label{eq:FTmutilde} 
\end{equation}

\end{widetext}

In Eqs.~\eqref{eq:FTTDGLE}-\eqref{eq:FTmutilde} we explicitly separated the quantity $A_1$, since it is the dominant Fourier component in the vicinity of the critical point $\jext=\jdp$ and plays an important role throughout the paper. Fourier coefficients with $|k|\gg 1$ quickly decay.
All other coefficients in this region can be found in the framework of perturbation theory.

Eq.~\eqref{eq:FTTDGLE} enables us to obtain the \IV characteristics in the complete domain of dynamical resistive states.  
The above mentioned minimization with respect to $E$ allows us to finds the shape of the \IV characteristics, which turns out to be critically dependent on the value of the dynamic coefficient $\u$ of the TDGLE. 

\begin{figure*}
\includegraphics[width=0.95\textwidth]{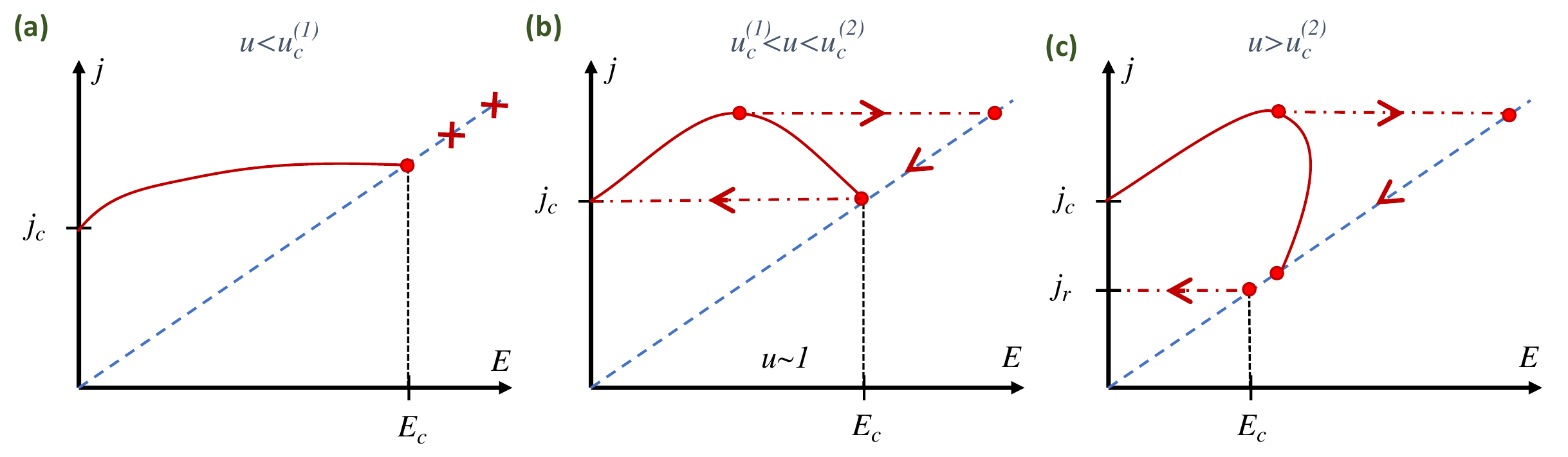}
\caption{Illustration of possible \IV (or $I$--$E$) curves for different $\u$. (a) shows the case of a reversible, non-hysteretic situation for small $\u$ below some value $\u_c^{(1)}$. Actual ``trajectories'' are indicated by arrow heads/crosses. (b) shows some intermediate $\u$ regime, where the realized \IV hysteresis has a reentrance current equal to the critical current $\jc$. (c) above a value $\u_c^{(2)}$ the reentrance current is lower than the critical current $\jc$. The latter is the realized hystersis curve.}\label{fig:IV}
\end{figure*}

One can expect to find three qualitatively different types of \IV characteristics, which are illustrated in Fig.~\ref{fig:IV}. 
The first one (Fig.~\ref{fig:IV}a) is reversible and could be realized when the dynamic coefficient $\u$ is sufficiently small:
$\u<\u_{c}^{(1)}$, where $\u_{c}^{(1)}$  is the first critical value, which will be obtained below.
In case  $\u>\u_{c}^{(1)}$ the \IV characteristics becomes irreversible (Fig.~\ref{fig:IV}b,c). 

In the case of strong damping, when $\u$ exceeds the second critical value $\u_{c}^{(2)}$, the transition to a finite value of the order parameter happens at a smaller $E_{c}$ value, which is where we define the \textit{reentrance} current $j_r$.
Calculation of the order parameter value in the vicinity of the critical point $(\u E)_{c}$ (see below) shows that in practice only the latter scenario of the \IV characteristics, shown in Fig.~\ref{fig:IV}c), is realized. 

\subsection{Current density close to the depairing current}
\begin{figure}[tbh]
\includegraphics[width=0.95\columnwidth]{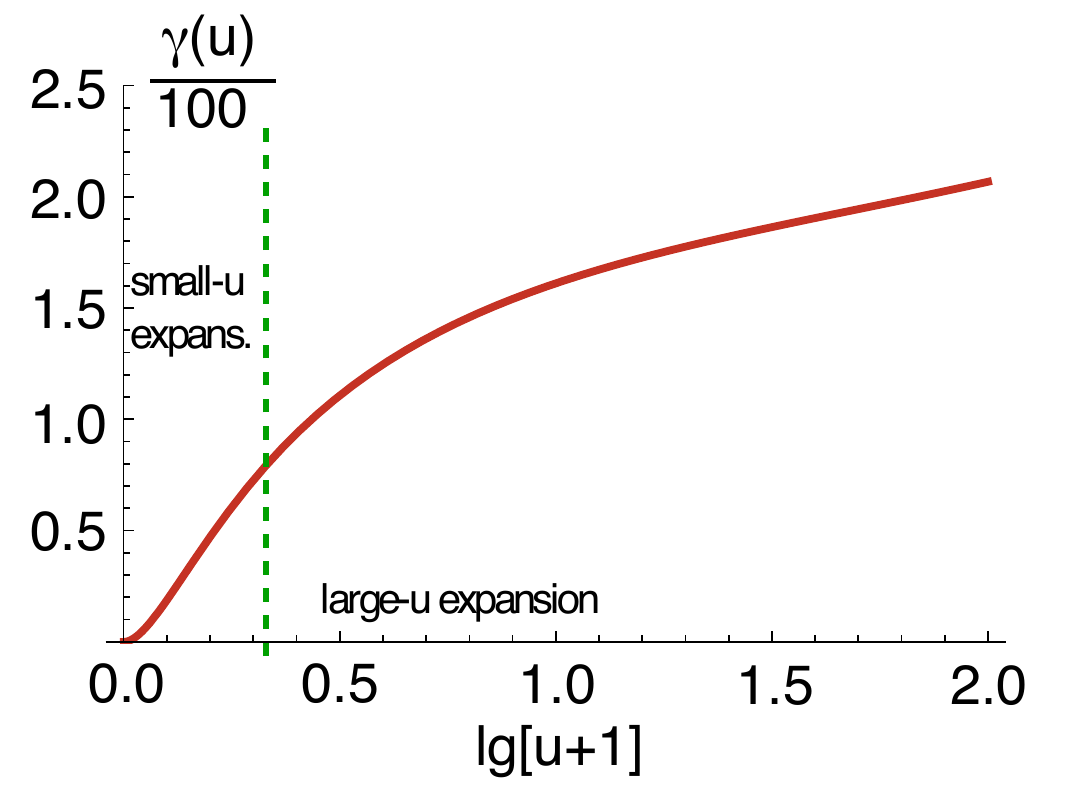}
\caption{Half-exponential plot of the coefficient $\gamma(\u)$ [Eq.~\eqref{eq:gammau}] of the $E^2$-term in our approximation for the current. Below $\u\sim 1$ (vertical dashed line) the small-$\u$ approximation   [Eq.~\eqref{eq:gammasmallu}] is indistinguishable from the exact curve, while for  $\u\gtrsim 1$ the large-$\u$ expression [Eq.~\eqref{eq:gammalargeu}] is indistinguishable from the exact one.  }
\label{fig:gamma}
\end{figure}

Next, we consider the case when the bias current density $\jext$ is close to $\jdp$. 
Eq.~\eqref{eq:TDGLE} allows us to obtain the above mentioned value of $\jdp$, which destroys superconductivity in the 1D channel, where the corresponding critical value of the order parameter is $\psi_{c}=\sqrt{2/3}$, and the value of wave-vector (for $E=0$) $Q_{c}=1/\sqrt{3}$.  
Close to this point,  Eq.~\eqref{eq:FTTDGLE} is decomposed into two equations with $\left\{ k,2-k\right\} $. 
A detailed analysis of the Fourier coefficients and the determination of the wave vector $Q$ is presented in Appendix~\ref{app:atjdp}. 
As a result of these calculations we obtain the electric field dependence of the current density to second order, close to its depairing value as
\begin{equation}
\jext=\jdp+E-E^2\gamma(\u)\,,\label{eq:japprox}
\end{equation}
where the calculation of the coefficient $\gamma(\u)$  of the quadratic term as function of $\u$ is a highly involved task, but can be performed exactly for any value of $\u$ and be expressed as a full derivative: 
\begin{widetext}
\begin{equation}
\gamma(\u)=
-3\sqrt{3}\u^2\frac{\partial}{\partial \u}\left[\frac{\pi}{2\sqrt{2\u}}\left\{\coth(\pi\sqrt{2\u})-\frac{1}{\pi\sqrt{2\u}}\right\}\left(1-\frac{6}{\u}+\frac{12}{\u^2}\right)+\frac{\pi^2}{\u}\left(1-\frac{2}{\u}\right)+\frac{4\pi^4}{15\u}\right]\,.\label{eq:gammau}
\end{equation}
The explicit expression of the full derivative is given in Appendix~\ref{app:atjdp}, Eq.~\eqref{eq:gammaexact}, and $\mathcal D$ is defined there in Eqs.~\eqref{eq:D} and~\eqref{eq:Dapprox}. 
Note, that the difference in braces behaves as $\pi\sqrt{2\u}/3$ for small $\u$, such that all terms $\propto \u^{-2}$ under the derivative cancel and the complete expression is non-singular at $\u=0$.
Therefore, in the limit of small $\u$ we keep the first two terms of the sum in~\eqref{eq:gammau} and expand the remaining sum to first order.  
This gives
\begin{equation}
\gamma(\u)=\frac{18}{\sqrt{3}} \u^2 \left(\frac{61}{(1 + 2 \u)^2 }+ \frac{7}{(4 + 2 \u)^2} +0.0486531-0.0185438 \u\right)\,.\label{eq:gammasmallu}
\end{equation}
For large values of $\u$ ($\u\gg 1$), we obtain from Eq.~\eqref{eq:gammau} (using the asymptotic expressions of the $\coth$ and $\sinh^{-2}$ terms)
\begin{equation}
\gamma(\u)= \sqrt{3}\left(3 \pi ^2+\frac{4\pi^4}{5}-\frac{3}{4}+\frac{3\pi}{8}\sqrt{2\u}\right)
-\frac{27\pi}{2} \sqrt{\frac{3}{2\u}} -\frac{3 \left(\sqrt{3} \left(4 \pi ^2-3\right)\right)}{\u}+\sqrt{\frac{3}{2}}  \frac{45\pi }{\u^{3/2}}-\frac{27 \sqrt{3}}{\u^2}+\mathcal{O}\left(\u e^{-2\pi\sqrt{2\u}}\right)\,.\label{eq:gammalargeu}
\end{equation}
The $\u$-dependence of $\gamma(\u)$ and the domains of validity of its approximations (\ref{eq:gammasmallu})-(\ref{eq:gammalargeu}) are presented in Fig.~\ref{fig:gamma}. $\u\sim 1$ separates the regions where small-$\u$ and large-$\u$ approximations work best, i.e., the relative deviations from the exact curve are both minimum at $\u=1.061$, less than $10^{-3}$.
\end{widetext}

\subsection{Vicinity of the critical points}
We now consider the vicinity of critical points, $(\u E)_{c}$, which are defined by the condition
\begin{equation}
\frac{\partial (\u E)}{\partial Q}=0\label{eq:uEc}\,.
\end{equation}
In this region we search for the solution to the non-linear problem~\eqref{eq:FTTDGLE} by its linear expansion over eigenfunctions.
Therefore the linearized condition of Eq.~\eqref{eq:uEc} can be understood as the following eigenvalue problem
\begin{equation}
\hat{L}\mathbf{f}=0\label{eq:Leq}\,,
\end{equation}
where the form of the linear operator $\hat{L}$ follows from Eq.~\eqref{eq:FTTDGLE}:
\begin{equation}
\hat{L}_{k,l}=
\begin{cases}
Z_{k-1}\delta_{l,1}+Z_{k-l}(1-\delta_{l,1})+(Q^2k^2-1)\delta_{k,l}&,\,\,k\neq 1\\
Z_{1-l}-(1-Q^2)\delta_{l,1}&,\,\,k= 1
\end{cases}
\end{equation}
with 
\begin{equation}
Z_k=
\begin{cases}
\frac{\u E}{kQ}(-1)^k&,\,\,k\neq 0\\
0&,\,\,k= 0
\end{cases}\,.
\label{eq:Z}
\end{equation}
The eigenvector $\mathbf{f}$ of Eq.~\eqref{eq:Leq} is related to the Fourier coefficients at the critical points in the following way
\begin{equation}\label{eq:EVdef}
\{f_k\}\leftrightarrow\{A_{k\geq 2},A_1,A_{k\leq 0}\}\,,
\end{equation}
where $f_1\leftrightarrow A_1$.

\begin{figure}
	\includegraphics[width=0.95\columnwidth]{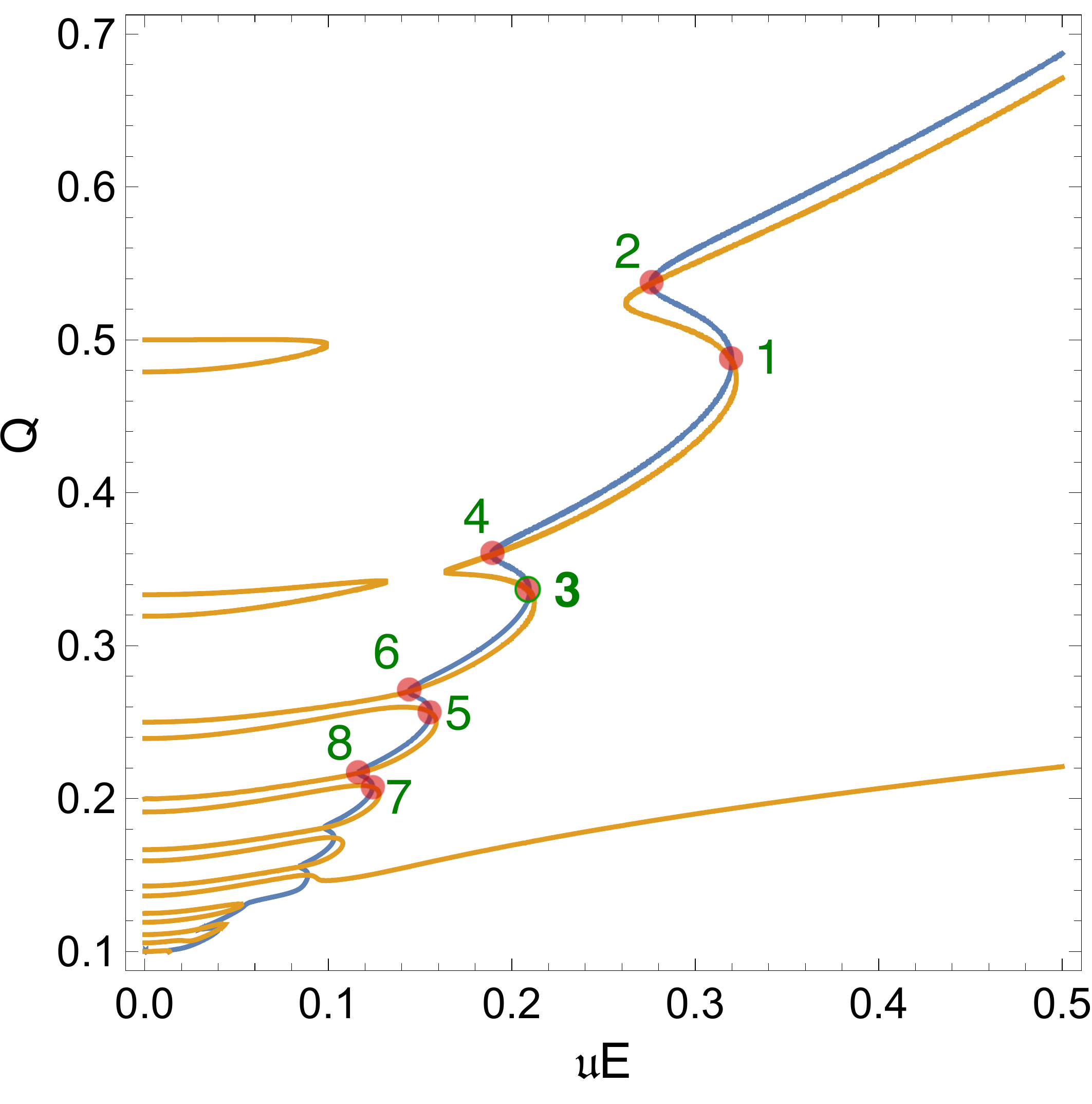}
	\caption{Solutions of EV problem for polynomial equations of order $Q^{70}$ and $(\u E)^{24}$ for $N_k=24$ Fourier components. Depicted are the ``zero'' lines of the polynomials related to $\det(\hat{L})$  (blue) and $\partial_Q\det(\hat{L})$ (orange) and intersection points corresponding to simultaneous eigenvalue pairs $\{(\u E)_{c},Q_c\}$ of $\hat{L}$ and $\hat{L}^\dagger$. The largest 8 are marked by circles and labeled.}
	\label{fig:poly}
\end{figure}

\begin{table}
	\centering
	\begin{tabular}{cccccc}
		\hline
		\hline
		$\nu$ & $(\u E)_{c}$ & $Q_c$ & $\lambda_1$ & $\lambda_2$ & $\lambda_3$\\
		\hline
		1 & 0.320 & 0.486 & 0.166 & -0.503 & -0.176\\
		2 & 0.276 & 0.537 & 0.102 & -0.230 & -0.0787\\
		\textbf{3} & \textbf{0.21} & \textbf{0.335} & \textbf{-0.0478} & \textbf{0.128} & \textbf{0.030}\\
		4 & 0.189 &  0.359 & 0.0265 & -0.0671 & -0.0169 \\
		5 & 0.156 & 0.256 & 0.0133 & -0.0345 & -0.00570 \\
		6 & 0.144 & 0.27 & 0.0074 & -0.0183 & -0.00258 \\
		7 & 0.124 & 0.208 & 0.00348 & -0.00922 & -0.000833 \\ 
		8 &0.116 & 0.217 & 0.00199 & -0.00528 & -0.000436 \\
		\hline
		\hline
	\end{tabular}
	\caption{\label{tab:EV24} The 8 largest simultaneous eigenvalue pairs $\{(\u E)_{c},Q_c\}$ of $\hat{L}$ and $\hat{L}^\dagger$ for $N_k=24$ Fourier components and corresponding solvability coefficients $\{\lambda_1,\lambda_2,\lambda_3\}$. Labels $\nu$ correspond to the intersection points shown in Fig.~\ref{fig:poly}.}
\end{table}

The above eigenvalue problem has possibly an infinite number of solutions $\{(\u E)_{c},Q_{c}\}$.
Note, that the quantities $\u$ and $E$ only appear in the form of a product at the critical point [in contrast to Eqs.~\eqref{eq:japprox} \& \eqref{eq:gammau}]. 
The linearized equation can be solved numerically and the largest eight critical points (defined by simultaneous eigenvalues of $\hat{L}$ and $\hat{L}^\dagger$)  are listed in table~\ref{tab:EV24} [the corresponding  normalized  eigenvectors of $\hat{L}$ and $\hat{L}^\dagger$ (here $\hat{L}^\dagger=\hat{L}^{\mathsf{T}}\neq\hat{L}$)  are given in the supplementary information]. Below we discuss the numerical solution in more detail. 

In order to get further insight into the behavior of the \IV characteristic near these critical points and to determine the type of instability point (first or second order transition), we introduce the operator $\delta\hat{L}$ as
\begin{equation}\label{eq:deltaL}
\delta\hat{L}_{k,l}=\delta Z_{k-1}\delta_{l,1}+\delta Z_{k-l}(1-\delta_{l,1})
\end{equation}
with
\begin{equation}
\delta Z_{k}=
\begin{cases}
\frac{\u\left(E-E_{c}\right)}{kQ_{c}}\left(-1\right)^{k} &,\,\, k\neq 0\\
0 &,\,\, k=0
\end{cases}
\end{equation}
[compare to Eq.~\eqref{eq:Z}].

Near a critical point $\left(\u E\right)_{c}$ the solution for the Fourier components in Eq.~\eqref{eq:FTTDGLE}  in our linearized approximation can be written in the form
\begin{equation}
\lambda(\u,E)f_{k}\,,
\end{equation}
with the proper permutation of $k$-indices as defined in Eq.~\eqref{eq:EVdef} and where the coefficient $\lambda(\u,E)$ follows from  the solvability condition
\begin{subequations}
\begin{eqnarray}
&& |\lambda(\u,E)|^{2}\left(\u \lambda_1+\lambda_2\right)+\sum_{k,k_1}\widetilde{f}_{k}^{*}\left(\delta\hat{L}_{k,k_{1}}\right)f_{k_{1}}=0\,, \label{eq:solv1}\\
&& \lambda_3 \u (E-E_c) = \sum_{k,k_1} \tilde{f^*_k}(\delta \hat L_{k,k_1}) f_{k_1}\,, \label{eq:solv2}
\end{eqnarray}
\end{subequations}
see  Appendix~\ref{app:uEc} for explicit expressions for all coefficients $\lambda_i$.

The expression for current density then takes the form 
\begin{equation}
j=E+|\lambda(\u,E)|^{2}\lambda_4 Q\,.
\end{equation}
(see Appendix~\ref{app:uEc} for definition of $\lambda_4$.)

We note, that in the critical region only the coefficient of the zero mode, $\lambda(\u,E)$, is dominant and
all other coefficients are small, scaling with $|(\u E)_{c}-\u E|/(\u E)_{c}$.

\begin{figure}
	\includegraphics[width=0.95\columnwidth]{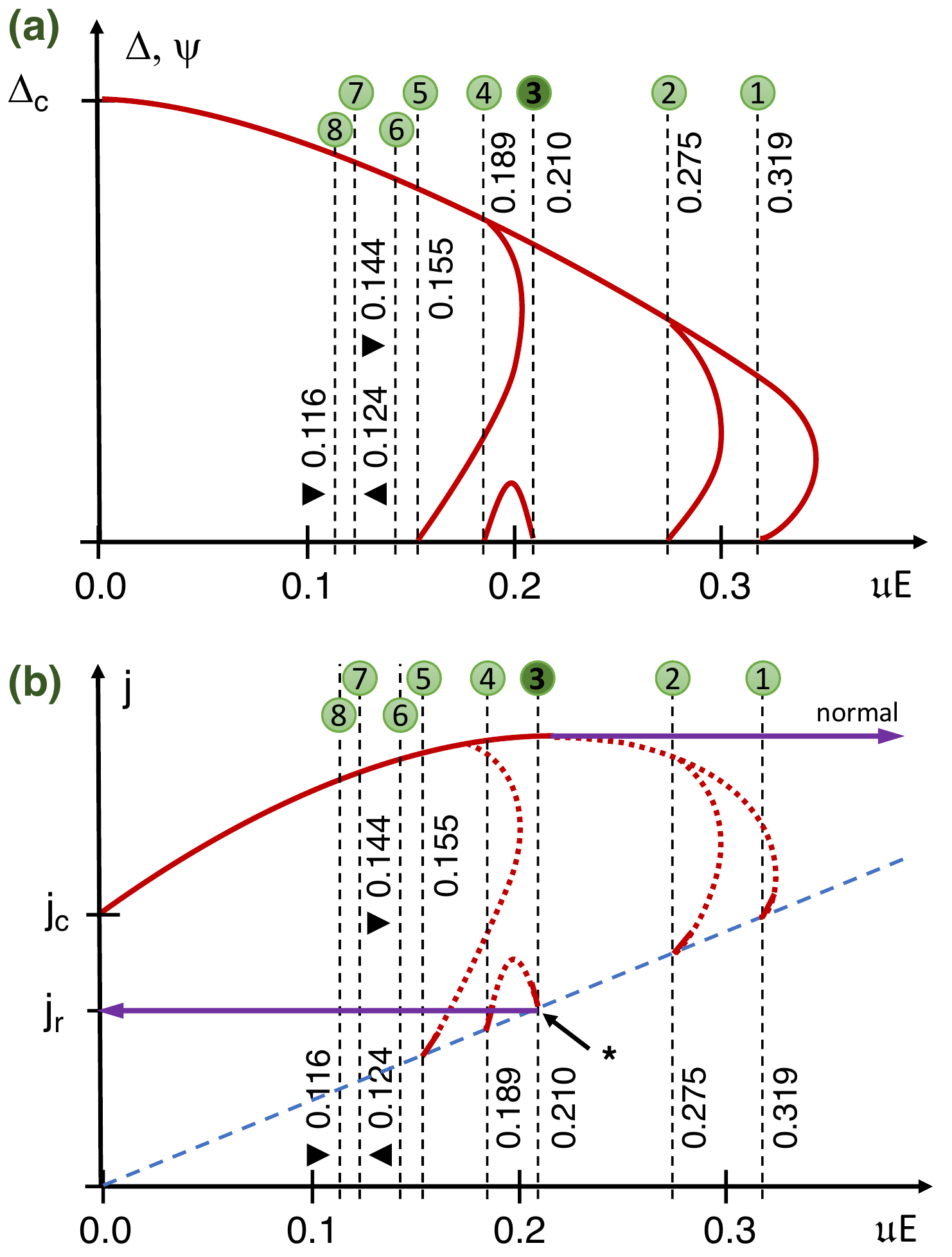}
	\caption{Critical behavior of the (a) superconducting order parameter and (b) the \IV characteristics, where the 8 largest eigenvalues of operators $\hat{L}$ and $\hat{L}^{\dagger}$ determine the intersection points of dynamical state and the linear Ohmic behavior, respectively. Compare to Fig.~\ref{fig:poly} and Table~\ref{tab:EV24}, values are indicated by vertical dashed lines and numbered accordingly. $\ast$ indicates the transition to superconducting state without threshold for critical value \raisebox{.5pt}{\textcircled{\raisebox{-.9pt} {3}}}. The slope of the \IV curves at the critical points is determined by Eq.~\eqref{eq:solv2}, indicated by solid line pieces. The connection of the critical points to the envelop is indicated by dotted lines (cannot be realized physically). Starting in the superconducting state at zero current, one follows up to $\jc$ upon increasing  $\jext$, where the system becomes resistive and eventually jumps into to normal state (indicated by an arrow) following the normal \IV curve (blue, dashed). When decreasing $\jext$ from the normal state, one follows down the normal \IV-line till the critical point $\ast$, where the slope of \IV is negative, such that fluctuating superconducting regions can grow and one jumps back into the superconducting state at $\jr$ (indicated by an arrow). At all other critical points we cannot follow the critical \IV (without threshold) as the slope is positive.}
	\label{fig:psiIV}
\end{figure}

Altogether, we can now analyze the critical point in detail. Therefore we calculate the  critical points $\{(\u E)_{c},Q_{c}\}$ numerically for truncated Fourier series with $N_k$ components (indices $k\in\{-N_k/2+1,\ldots, N_k/2\}$). These are obtained as simultaneous solutions of the polynomial equations $Q^{N_k}\det(\hat{L})=0$ and $Q^{N_k+1}\partial_Q\det(\hat{L})=0$ of order $3N_k-2$ in $Q$.
Figure~\ref{fig:poly} shows the solutions for $N_k=24$ (order $Q^{70}$), where the solid lines represent the solutions of the individual equations.
Note, that solving the linearized equations for the truncated Fourier series  leaves the largest critical values $\{(\u E)_{c},Q_{c}\}$ invariant for sufficiently large $N_k$.
For these solutions we can then obtain the eigenvectors of $\hat{L}$ and $\hat{L}^\dagger$, which allows us then to extract the behavior of the \IV characteristic near these critical points by evaluation of the parameters $\lambda_1$, $\lambda_2$, and $\lambda_3$. At those points new branches appear, which can bring the system out of the normal state.
Using Eq.~\eqref{eq:solv2} then defines the slope of the linearized \IV characteristic. The numerical calculation reveals that the critical value $\nu=3$, $(\u E)_c^{(3)} \approx 0.21$, has locally a negative slope among the eight largest $(\u E)_c$ values, indicating that a reentrance into the superconducting state can happen without a threshold (second order) at a current density $\jr$. Fig.~\ref{fig:psiIV} shows the behavior of the order parameter $\Delta$, (a), and current $\jext$, (b), at the critical points as a function of $\u E$. The connection of the critical points to the envelop (defined by the \IV curve for increasing current) is indicated by dotted lines (these cannot be realized physically). 
Practically, one can make a `hysteresis `loop'' in the $\jext$-$(\u E)$ diagram by starting in the superconducting state at zero current, following up to $\jc$ upon increasing  $\jext$, where the system becomes resistive and eventually jumps into to normal state (indicated by an arrow) following the normal \IV curve (blue, dashed). 
When decreasing $\jext$ from the normal state, one follows down the normal \IV-line till the critical point labeled $\ast$ ($\nu=3$), where the slope of \IV is negative, such that fluctuating superconducting regions can grow and one jumps back into the superconducting state at $\jr$ (indicated by an arrow). Below this point the normal state is always unstable.
At all other (larger) critical points we cannot follow the critical \IV (without threshold) as the slope is positive (first order).
We note that the specific picture depends on the actual value of $\u$ for the physical system under consideration, here we assume a value of $\u$ of order one.

The numerical analysis demonstrates that probably an infinite set of solutions of the eigenvalue problem, Eq.~\eqref{eq:Leq}  exists.

In analogy to finding the (global) extremum of a function on a finite support, where also the boundary values need to be checked, here 
 we should also study the properties of the system close to the hypothetical ``endpoints'', if those exist [besides the (local) critical points defined by~\eqref{eq:Leq}]. 
By ``endpoints'' we mean  points of the surface $\left\{ \hat{L}\psi=0\right\} $ in Hilbert space, where the value $\left(\u E\right)$ reaches its maxima under the condition 
\begin{equation}
\sum_{k=2}^{\infty}k|A_{k}|^{2}+|A_1|^{2}-\sum_{k=1}^{\infty}k|A_{-k}|^{2}=0.
\end{equation}
Our numerical evaluation of this condition reveal that such ``endpoints'' are irrelevant.

\section{Conclusions} \label{sec:conclusions}

We have investigated the \IV characteristic of a superconducting strip in the region near the depairing current $\jdp$ and studied the instability points of the normal state as function of the current.
Interestingly, one finds a degeneracy in second order perturbation theory in the electric field $E$ by solving  linearized equation at those critical points.

This degeneracy leads to the appearance of additional branches splitting off from the Ohmic behavior seen in Fig.~\ref{fig:psiIV}.
Numerically, we calculated the critical points and found that the largest electric field value, at which the normal state first becomes unstable upon decreasing the current, i.e., indicating the possibility of a transition into the superconducting state with finite value of the order parameter amplitude (see Fig.~\ref{fig:psiIV}a), and a branch in the \IV appears is at $\u E=0.3199$. 
However, the slope of the branch is positive, indicating a first order transition, which we can only follow when increasing the current (and eventually jumping back in the normal state).
Therefore, when increasing the current for fixed $\u E$ in the intervals $[0.21,0.275[$ or $]0.275,0.319]$, a transition into the superconducting state will happen at $\u E=0.275$ or $\u E=0.319$, respectively, before returning to the normal state at larger currents. 

However, most importantly, we also found the smallest value for $\u E$ when the normal state is always unstable to be equal to $0.2095$ (indicated by $\ast$ in Fig.~\ref{fig:psiIV}), defining the reentrance current into the superconducting state.
In contrast to the evaluation of the critical current, the evaluation of the re-entrance current is significantly more involved.

\section*{Acknowledgments} 
We are delighted to thank I.S. Aranson for interesting discussions. 
The research was supported by the U.S. Department of Energy, Office of Science, Basic Energy Sciences, Materials Sciences and Engineering Division.

Yu.N.O acknowledges Prof. Dr. Jeroen van den Brink for his hospitality in LIFW and the DFG for granting him a Mercator-Professoren-Stipendium.
A. V. acknowledges financial support from the project CoExAn (HORIZON 2020, grant agreement 644076), from Italian MIUR through the PRIN 2015 program (Contract No. 2015C5SEJJ001).

\appendix

\section{Solution of TDGLE near the depairing current}\label{app:atjdp}

 Close to the depairing current,  Eq.~\eqref{eq:FTTDGLE} is decomposed into pairwise equations with $\left\{ k,-k+2\right\} $.
\begin{widetext}
	For $k\neq$1 
	\[
	\left(\begin{array}{cc}
	\left[1-2|A_1|^{2}-\frac{\u|A_1|^{2}\left(k+1\right)}{2\left(k-1\right)}-k^{2}Q^{2}\right] & -A_1^{2}\left(1+\frac{u\left(3-k\right)}{2\left(k-1\right)}\right)\\
	-\left(A_1^{*}\right)^{2}\left(1-\frac{u\left(k+1\right)}{2\left(k-1\right)}\right) & \left[1-2|A_1|^{2}+\frac{u|A_1|^{2}\left(3-k\right)}{2\left(k-1\right)}-\left(k-2\right)^{2}Q^{2}\right]
	\end{array}\right)\left(\begin{array}{c}
	A_{k}\\
	A^*_{-k+2}
	\end{array}\right)=\frac{\u E\left(-1\right)^{k}}{\left(k-1\right)Q}\left(\begin{array}{c}
	-A_1\\
	A_1^{*}
	\end{array}\right)
	\]
	
	Solving this system results in 
	\begin{subequations}
		\begin{equation}
		A_{k}=-\frac{1}{\D}\frac{\u E\left(-1\right)^{k}A_1}{\left(k-1\right)Q}\left[1-3|A_1|^{2}-\left(k-2\right)^{2}Q^{2}\right],\qquad A^*_{-k+2}=\frac{1}{\D}\frac{\u E\left(-1\right)^{k}A_1^{*}}{\left(k-1\right)Q}\left[1-3|A_1|^{2}-k^{2}Q^{2}\right]\,,\label{eq:coeffA}
		\end{equation}
		where 
		\begin{equation}
		\D=\left[1-2|A_1|^{2}-\left(\left(k-1\right)^{2}+1\right)Q^{2}-\frac{\u|A_1|^{2}}{2}\right]^{2}-\left[2\left(k-1\right)Q^{2}+\frac{\u|A_1|^{2}}{k-1}\right]^{2}-|A_1|^{4}\left[\left(1-\frac{\u}{2}\right)^{2}-\frac{\u^{2}}{\left(k-1\right)^{2}}\right].\label{eq:D}
		\end{equation}
	\end{subequations}
\end{widetext}

In our approximation, we obtain from Eq.~\eqref{eq:D} 
\begin{equation}
\D=\frac{1}{9}(k-1)^{2}((k-1)^{2}+2\u)\label{eq:Dapprox}
\end{equation}
when the electric field $E$ is much smaller than its critical value $E_{c}$.

For $k=1$, we obtain from Eq.~\eqref{eq:FTTDGLE}, the following equation for the quantity $|A_1|^2$ in second order perturbation theory
\begin{widetext}
	\begin{eqnarray}
	&&\sum_{k=2}^\infty\frac{\u }{k-1}\left\{\frac{ E (-1)^k}{Q} (A_k-A_{2-k})-\frac{A_1}{2}\left((k+1)|A_k|^2-(3-k)|A_{k-2}|^2\right)+(k-1)A_1^*A_kA_{2-k}\right\}=\nonumber\\
	&&A_1(1-|A_1|^2-Q^2)-2\sum_{k=2}^\infty\left\{A_1(|A_k|^2+|A_{2-k}|^2)+A_1^*A_kA_{2-k}\right\}\label{eq:PT2}
	\end{eqnarray}
	
	Using second order perturbation theory, we can set
	\begin{equation}
	|A_1|=\sqrt{\frac{2}{3}}+\beta E^2\,,\quad Q=\frac{1}{\sqrt{3}}+\alpha E^2
	\label{eq:albe}
	\end{equation}
	where $\{\alpha,\beta\}$ are constants.
	Inserting expression~\eqref{eq:coeffA} for the coefficients $A_k$ and expressions~\eqref{eq:albe} into Eq.~\eqref{eq:PT2}, we get the following relation for those constants:
	\begin{equation}
	\frac{2}{\sqrt{3}}(\alpha+\sqrt{2}\beta)=\sum_{k=2}^\infty\left\{\left(\frac{4\u^3}{3\D^2}-\frac{6\u^2}{(k-1)^2\D}\right) \left(1+\frac{(k-1)^2+1}{3}\right)-\frac{4\u^2}{\D^2}\left[\frac{1}{(k-1)^2}\left(1+\frac{(k-1)^2+1}{3}\right)^2+\frac{4}{3}\right]\right\}\label{eq:ab}
	\end{equation}
	
One important property of Eqs.~\eqref{eq:PT2} and~\eqref{eq:albe} should be mentioned: in second order perturbation theory, defined by Eq.~\eqref{eq:ab}, $\alpha$ and $\beta$ appear only in  combination $\alpha+\sqrt2 \beta$. This implies that corrections to the quantities $|A_1|$ and $Q$  appear separately in  perturbation theory only in order $\mathcal{O}\left(E^{4}\right)$
	
In the same approximation we obtain from Eq.~\eqref{eq:jansatz}
	\begin{equation}
	j=\frac{2}{3\sqrt{3}}+E+\frac{2}{3}E^2(\alpha+\sqrt{2}\beta)+\frac{4\u^2E^2}{\sqrt{3}}\sum\limits_{k=0}^{\infty}\frac{1}{(k-1)^2\D^2}\left[\left(1+\frac{(k-1)^2+1}{3}\right)^2-\frac{4}{3}(k-1)^2-\frac{4}{9}(k-1)^4\right]\,.\label{eq:jappprox}
	\end{equation}
	Inserting Eq.~\eqref{eq:ab} into expression~\eqref{eq:jappprox} yields an expression for the current $j$ in second order  perturbation theory in $E$:
	\begin{subequations}
	\begin{eqnarray}
	j&=&\frac{2}{3\sqrt{3}}+E-E^2\gamma(\u)\label{eq:j2ndorder}\\
	\gamma(\u)&=&\frac{2\u^2}{\sqrt{9}}\sum\limits_{k=0}^{\infty}\frac{1}{\D^2}\left[(k-1)^4+12(k-1)^2+48\right]\label{eq:gamma2ndo}
	\end{eqnarray}
\end{subequations}
	
	The  expression for $\gamma(\u)$ in~\eqref{eq:gamma2ndo} can be evaluated explicitly as
	\begin{eqnarray}
	\gamma(\u)=\frac{\sqrt{3} }{40 \u^2}&&\left[ 32 \pi ^4 \u^2-30 (\u-6)^2+15 \pi  \sqrt{2} ((\u-18) \u+60) \sqrt{\u} \coth \left( \pi  \sqrt{2\u}\right)+\right.\nonumber\\
	&&\,\,\left.30 \pi ^2 \u \left(4 (\u-4)+((\u-6) \u+12) \sinh^{-2}\left( \pi  \sqrt{2\u}\right)\right)\right]\,,\label{eq:gammaexact}
	\end{eqnarray}
	which can be written as the full derivative Eq.~\eqref{eq:gammau}.

\end{widetext}	

\section{Solvability and \IV at the critical points}\label{app:uEc}

The coefficients in the solvability equations, \eqref{eq:solv1} and \eqref{eq:solv2}, are given by

\begin{widetext}
\begin{subequations}
	\begin{eqnarray}
	\lambda_1&= & \frac{1}{2}\sum_{k}\widetilde{f}_{k}^{*}\left\{ \sum_{l\neq 0}\frac{1}{l}\left[\left(f_{1}^{2}\left(2-l\right)f_{-l+1}^{*}+|f_{1}|^{2}\left(2+l\right)f_{l+1}\right)\delta_{k,l+1}+ \right.\right.\nonumber\\
	&&\left.\left.\sum_{m\neq 1}\left(f_{1}\left(2-l\right)f_{-l+1}^{*}f_{m}+f_{1}^{*}\left(2+l\right)f_{l+1}f_{m}\right)\delta_{k,l+m}\right]\right.\nonumber\\
	&& -\frac{1}{2}\sum_{{l,m\neq 1} \atop {l\neq m}}\frac{l+m}{l-m}\left(f_{1}f_{l}^{*}f_{m}\delta_{k,-l+m+1}+\sum_{n\neq 1}f_{l}^{*}f_{m}f_{n}\delta_{k,m+n-l}\right)\,,\label{eq:lambda1}\\
	\lambda_2&= & \sum_{k}\widetilde{f}_{k}^{*}\left\{ f_{1}|f_{1}|^{2}\delta_{k,1}+2|f_{1}|^{2}\left(1-\delta_{k,1}\right)f_{k}+\right.\nonumber\\
	&&\left.\sum_{l\neq 1}\left[2\sum_{m\neq 1}\left(f_{1}f_{l}f_{m}^{*}\delta_{k,l-m+1}+\sum_{n\neq 1}f_{l}f_{m}^{*}f_{n}\delta_{k,l-m+n}\right)+f_{1}^{2}f_{l}^{*}\delta_{k,2-l}+\sum_{m\neq 1}f_{1}^{*}f_{l}f_{m}\delta_{k,l+m-1}\right]\right\} \,,\label{eq:lambda2}\\	
	\lambda_3&=&\frac{1}{Q_c}\left[\sum_{k\neq k_1} \frac{(-1)^{k-k_1}}{k-k_1}\widetilde{f}_k^*f_{k_1}\right]\,,\label{eq:lambda3} \\ 
	\lambda_{4}&=&\sum_{k=-\infty}^{\infty}k|f_{k}|^{2}\,.\label{eq:lambda4}
	\end{eqnarray}
\end{subequations}
\end{widetext}
Here $\widetilde{f}_{k}$ are the components of the normalized eigenvector of the transposed operator $\hat{L}^\dagger=\hat{L}^{\mathsf{T}}$.

Next, we define the function
\[
\Lambda(\u E)\equiv \sum_{k,k_1}\widetilde{f}_{k}^{*}\left(\delta\hat{L}_{k,k_{1}}\right)f_{k_{1}}=-|\lambda(\u,E)|^2(\u\lambda_1+\lambda_2)\,,
\]
where $f_k$ are the components of the eigenvector of $\hat{L}$ and $\widetilde{f}_{k}$ are the components of the normalized eigenvector of the operator $\hat{L}^\dagger$. Function $\delta\hat{L}_{k,k_{1}}$ is defined in Eq.~\eqref{eq:deltaL}.
Eq.~\eqref{eq:FTTDGLE} in the vicinity of each critical point $\left\{ \left(\u E\right)_{c},Q_{c}\right\}$ can then be rewritten in the form
\begin{equation}
|\lambda(\u,E)|^{2}\left(\u \lambda_1+\lambda_2\right)+\lambda_3\u\left(E-E_{c}\right)=0\,,
\end{equation}
where the coefficients are given in~\eqref{eq:lambda1}-\eqref{eq:lambda3}.
Therefore:
\[
|\lambda(\u,E)|^2=-\frac{\Lambda(\u E)}{\u\lambda_1+\lambda_2}\,.
\]
The functions $\Lambda(\u E)$ corresponding to the eight largest critical value $(\u E)_c$ are given by

\begin{eqnarray*}
	\Lambda^{(1)}(\u E)&=& 0.0564 - 0.176 \u E\,,\\
	\Lambda^{(2)}(\u E)&=&0.0217 - 0.0787  \u E\,,\\
	\mathbf{\Lambda^{(3)}(\u E)}&=&\mathbf{-0.00629 + 0.03  \u E}\,,\\
	\Lambda^{(4)}(\u E)&=&0.00319 - 0.0169  \u E\,,\\
	\Lambda^{(5)}(\u E)&=&0.000888 - 0.00570  \u E\,,\\
	\Lambda^{(6)}(\u E)&=&0.000372 - 0.00258  \u E\,,\\
	\Lambda^{(7)}(\u E)&=&0.000103 - 0.000833  \u E\,,\\
	\Lambda^{(8)}(\u E)&=&0.0000507 - 0.000436  \u E\,.
\end{eqnarray*}
Note the signs of the coefficients in $\Lambda^{(3)}$. For completeness, we reproduce the corresponding eigenvectors $f_k$  and $\widetilde{f}_{k}$ in the supplementary information.

\section{Derivation  of Fourier equations}
Here we will obtain the equation system for the Fourier coefficients $A_k$. From Eq.~\eqref{eq:FTTDGLE} we obtain for $k_0\neq 1$
\begin{widetext}
	\begin{eqnarray}
	Z_{k_0-1}A_1+\sum_{k\neq 1}Z_{k_0-k}A_k+\frac{\u}{2(k_0-1)}\left[A_1^2(3-k_0)A^*_{2-k_0}+|A_1|^2(k_0+1)A_{k_0}\right]+&&\nonumber\\
	\sum_{k\notin \{0,k_0-1\}}\frac{\u}{2k}\left[A_1(2-k)A_{k_0-k}A^*_{1-k}+(2+k)A_1^*A_{k+1}A_{k_0-k}\right]+&&\nonumber\\
	\frac{\u}{2}\sum_{k\neq \{1,2-k_0\}}\left[\frac{2k+k_0-1}{k_0-1}A_1A_k^*A_{k_0+k-1}-\sum_{q\notin \{1,k\}}\frac{k+q}{k-q}A_k^*A_qA_{k_0+k-q}\right]&=&\nonumber\\
	+(1-2|A_1|^2-Q^2k_0^2)A_{k_0}-2\sum_{k\notin \{1, k_0\}}A_1A_kA_{k-k_0+1}^*&&\nonumber\\
	-\sum_{k\neq 1,q\notin\{1,k-K_0+1\}}A_kA_q^*A_{k_0+q-k}-A_1^2A^*_{2-k_0}-A_1^*\sum_{k\notin \{0,k_0\}}A_kA_{k_0-k+1}\label{eq:A1}
	\end{eqnarray}
From Eq.~\eqref{eq:FTTDGLE} we obtain a separate equation for $k_0=1$:
	\begin{eqnarray}
	&&\sum_{k\neq 1}\left[Z_{1-k}A_k-\frac{\u(k+1)}{2(k-1)}\left(A_1|A_k|^2-A_1^*A_kA_{2-k}\right)-\frac{u}{2}\sum_{q\notin \{1,k\}}A_k^*A_qA_{k+1-q}\right]\nonumber\\
	&=&A_1(1-|A_1|^2-Q^2)-\sum_{k\neq 1}\left[2A_1|A_k|^2+A_1^*A_kA_{2-k}+\sum_{q\notin\{1,k\}}A_kA_q^*A_{1+q-k}\right]\label{eq:A2}
	\end{eqnarray}
Next, we introduce the operator $\hat{M}_{k_0}$ for $k_0\neq 1$ as
\begin{equation}
\hat{M}_{k_0}=\left(\begin{array}{cc} (1-2|A_1|^2-Q^2k_0^2)-\frac{\u|A_1|^2(k_0+1)}{2(k_0-1)}&-A_1^2\left(1+\frac{\u(3-k_0)}{2(k_0-1)}\right) \\ 
-(A_0^*)^2\left(1-\frac{\u(k_0+1)}{2(k_0-1)}\right) & 1-2|A_1|^2-Q^2(k_0-2)^2+\frac{\u|A_1|^2(3-k_0)}{2(k_0-1)}\end{array}\right)
\end{equation}

With this, Eq.~\eqref{eq:TDGLE} can be written in the form
\begin{equation}\label{eq:A4}
\hat{M}_{k_0}\left(\begin{array}{l} A_{k_0} \\ A^*_{2-k_0}\end{array}\right)=\left(\begin{array}{l} Z_{k_0-1}A_1+\sum_{k\neq 1}Z_{k_0-k}A_k \\ Z_{1-k_0}A_1^*+\sum_{k\neq 1}Z_{k-k_0}A_{2-k}^* \end{array}\right)+\left(\begin{array}{l} \Phi_{k_0}\\ \Phi^*_{2-k_0}\end{array}\right)
\end{equation}
where $\Phi_{k_0}$ is
\begin{eqnarray}
\Phi_{k_0}&=&\frac{\u}{2}\sum_{k\notin \{1,k_0\}}\left[\frac{2k+k_0-1}{k_0-1}A_1A_k^*A_{k_0+k-1}-\sum_{q\notin\{1,k\}}\frac{k+q}{k-q}A_k^*A_qA_{k_0+k-q}\right]\nonumber\\
&&+\sum_{k\notin \{1,k_0\}}\frac{k}{2(k-1)}\left[A_2(3-k)A^*_{2-k}+A^*_2(1+k)A_k\right]A_{k_0-k+1}\nonumber\\
&&+2\sum_{k\notin \{1,k_0\}}A_1A_kA^*_{k-k_0+1}+\sum_{k\neq 1,q\notin\{1,k-k_0+1\}}A_kA_q^*A_{k_0+q-k}+A_1^*\sum_{k\notin \{0,k_0\}}A_kA_{k_0-k+1}
\end{eqnarray}
\end{widetext}
Note, that a free parameter  $Q$ appears in Eqs.~\eqref{eq:A2} and~\eqref{eq:A4}.
The value  of this parameter is found by the extremal condition for the electric field $E$ for a given current density $j$.

\bibliographystyle{apsrev4-1}

\newpage
\onecolumngrid
\newpage

\section{Supplementary Information: Eigenvectors near critical points}

Here we present detailed results of the numerical evaluation at critical points including the eigenvectors.
All calculations are done for $N_k=24$ (Fourier) components of the eigenvectors of $\hat{L}$ and $\hat{L}^\dagger$,  $f_k$ and $\widetilde{f}_{k}$, respectively.
Here component indices range from $k=-N_k/2+1\ldots N_k/2$.

The results for the eight largest critical $(\u E)_c$ are reproduced in table~\ref{tab:EV24ex}, including the critical structural constant $Q_c$, and the four parameters $\lambda_i$ ($i=1,2,3,4$); see manuscript.
The bold printed critical value 3 corresponds to the reentrance point.

\begin{table}
	\centering
	\begin{tabular}{ccccccc}
		\hline
		\hline
		$\nu$ & $(\u E)_{c}$ & $Q_c$ & $\lambda_1$ & $\lambda_2$ & $\lambda_3$ & $\lambda_4$\\
		\hline
		1 & 0.320 & 0.486 & 0.166 & -0.503 & -0.176 & 2.07\\
		2 & 0.276 & 0.537 & 0.102 & -0.230 & -0.0787 & 1.97\\
		\textbf{3} & \textbf{0.21} & \textbf{0.335} & \textbf{-0.0478} & \textbf{0.128} & \textbf{0.030} & \textbf{2.93}\\
		4 & 0.189 &  0.359 & 0.0265 & -0.0671 & -0.0169 & 2.74\\
		5 & 0.156 & 0.256 & 0.0133 & -0.0345 & -0.00570 & 3.83\\
		6 & 0.144 & 0.27 & 0.0074 & -0.0183 & -0.00258 &3.63 \\
		7 & 0.124 & 0.208 & 0.00348 & -0.00922 & -0.000833 & 4.74\\ 
		8 &0.116 & 0.217 & 0.00199 & -0.00528 & -0.000436 & 4.54\\
		\hline
		\hline
	\end{tabular}
	\caption{\label{tab:EV24ex} Similar to table I in the paper: The 8 largest simultaneous eigenvalue pairs $\{(\u E)_{c},Q_c\}$ of $\hat{L}$ and $\hat{L}^\dagger$ for $N_k=24$ Fourier components and  constants $\{\lambda_1,\lambda_2,\lambda_3,\lambda_4\}$. The corresponding coefficients $\Lambda(\u E)$ are given with the values of the eigenvectors (see text).}
\end{table}

\begin{widetext}
	The normalize eigenvectors and functions $\Lambda(\u E)$ (see manuscript) are listed below, where bracketed superscripts correspond to the index $\nu$ in table~\ref{tab:EV24ex}. The $k=1$component is printed in bold.
	{
		\tiny
		\begin{eqnarray*}
			\Lambda^{(1)}(\u E)&=& 0.0564 - 0.176 \u E\\
			\mathbf{f}^{(1)}&=&\left\{
			0.000824, -0.00111, 0.00155, -0.00228, 0.00355, -0.00591, 0.0107, -0.0210, 0.0420, -0.0629, 0.000279, -0.124, \right.\\
			&&\left.\textbf{-0.216}, -0.871, -0.416, -0.0152, -0.00513, 0.00315, -0.00203, 0.00138, -0.000974, 0.000714, -0.000538, 0.000412\right\}\\
			\widetilde{\mathbf{f}}^{(1)}&=&\left\{
			0.000495, -0.000664, 0.000909, -0.00129, 0.00190, -0.00295, 0.00481, 0.0142, 0.390, 0.815, 0.404, 0.117, \right.\\
			&&\left.\textbf{ -0.000270}, 0.0589, -0.0394, 0.0196, -0.0100, 0.00553, -0.00332, 0.00213, -0.00145, 0.00103, -0.000758, 0.000580\right\}\\
			\Lambda^{(2)}(\u E)&=&0.0217 - 0.0787  \u E\\
			\mathbf{f}^{(2)}&=&\left\{
			-0.000251, 0.000327, -0.000441, 0.000616, -0.000898, 0.00138, -0.00226, 0.00396, -0.00698, 0.00708, 0.0265, 0.0704,\right.\\
			&&\left.\textbf{0.271}, 0.933, 0.226, -0.0119, 0.00560, -0.00280, 0.00160, -0.00101, 0.000672, -0.000472, 0.000343, -0.000256\right\}\\
			\widetilde{\mathbf{f}}^{(2)}&=&\left\{
			-0.000308, 0.000425, -0.000607, 0.000909, -0.00145, 0.00254, -0.00507, 0.0108, -0.204, -0.844, -0.490, -0.0638,\right.\\
			&&\left.\textbf{-0.0240}, -0.00643, 0.00633, -0.00359, 0.00205, -0.00125, 0.000813, -0.000557, 0.000399, -0.000295, 0.000225, -0.000177\right\}\\
			\Lambda^{(3)}(\u E)&=&-0.00629 + 0.03  \u E\\
			\mathbf{f}^{(3)}&=&\left\{
			0.000672, -0.000894, 0.00126, -0.00186, 0.00288, -0.00466, 0.00773, -0.0121, 0.0142, -0.00655, 0.0127, 0.0277, \right.\\
			&&\left.\textbf{0.0756}, 0.482, 0.766, 0.414, 0.0509, 0.00277, -0.00123, 0.000876, -0.00063, 0.000463, -0.000348, 0.000264\right\}\\
			\widetilde{\mathbf{f}}^{(3)}&=&\left\{
			-0.000334, 0.000453, -0.00062, 0.000864, -0.00121, 0.00275, 0.0504, 0.411, 0.76, 0.478, 0.15, 0.0275,\right.\\
			&&\left.\textbf{0.0126}, -0.00651, 0.0141, -0.012, 0.00766, -0.00461, 0.00284, -0.00183, 0.00124, -0.000873, 0.000641, -0.000493\right\}\\
			\Lambda^{(4)}(\u E)&=&0.00319 - 0.0169  \u E\\
			\mathbf{f}^{(4)}&=&\left\{
			0.00019, -0.000243, 0.000323, -0.000444, 0.000631, -0.000922, 0.00135, -0.00176, 0.00104, 0.00278, 0.00107, 0.0279,\right.\\
			&&\left.\textbf{0.0712}, 0.565, 0.778, 0.263, 0.00639, 0.00288, -0.00158, 0.000952, -0.000615, 0.000419, -0.000298, 0.000216\right\}\\
			\widetilde{\mathbf{f}}^{(4)}&=&\left\{
			0.000288, -0.000412, 0.000607, -0.000942, 0.00157, -0.00285, -0.00635, -0.261, -0.773, -0.561, -0.141, -0.0276,\right.\\
			&&\left.\textbf{-0.00108}, -0.00274, -0.00105, 0.00176, -0.00135, 0.000918, -0.000627, 0.00044, -0.000319, 0.000239, -0.000184, 0.000147\right\}\\
			\Lambda^{(5)}(\u E)&=&0.000888 - 0.00570  \u E\\
			\mathbf{f}^{(5)}&=&\left\{
			-0.000476, 0.000618, -0.000857, 0.00124, -0.00182, 0.00265, -0.00356, 0.00375, -0.00254, 0.00251, 0.000479, 0.0111,\right.\\
			&&\left.\textbf{0.0244}, 0.201, 0.522, 0.714, 0.412, 0.0800, 0.00481, -0.000369, 0.000328, -0.000249, 0.000189, -0.000142\right\}\\
			\widetilde{\mathbf{f}}^{(5)}&=&\left\{
			-0.000179, 0.000243, -0.000323, 0.000365, -0.00480, -0.0800, -0.412, -0.714, -0.522, -0.200, -0.0488, -0.0111,\right.\\
			&&\left.\textbf{-0.000473}, -0.00251, 0.00255, -0.00375, 0.00356, -0.00265, 0.00181, -0.00123, 0.000845, -0.000601, 0.000444, -0.000348\right\}\\
		\end{eqnarray*}
		
		\begin{eqnarray*}
			\Lambda^{(6)}(\u E)&=&0.000372 - 0.00258  \u E\\
			\mathbf{f}^{(6)}&=&\left\{
			0.000121, -0.000149, 0.000193, -0.000254, 0.000336, -0.000424, 0.000445, -0.00019, -0.000425, 0.000347, -0.00197, -0.00632,\right.\\
			&&\left.\textbf{-0.0236}, -0.216, -0.598, -0.714, -0.29, -0.0276, -0.00173, 0.000723, -0.000465, 0.00031, -0.000215, 0.000151\right\}\\
			\widetilde{\mathbf{f}}^{(6)}&=&\left\{
			0.000206, -0.000305, 0.00046, -0.000719, 0.00172, 0.0276, 0.29, 0.714, 0.597, 0.216, 0.0471, 0.00632,\right.\\
			&&\left.\textbf{0.00196}, -0.000335, 0.000408, 0.000212, -0.000461, 0.000434, -0.000341, 0.000256, -0.000193, 0.000148, -0.000116, 0.0000949\right\}\\
			\Lambda^{(7)}(\u E)&=&0.000103 - 0.000833  \u E\\
			\mathbf{f}^{(7)}&=&\left\{
			-0.000306, 0.000374, -0.000496, 0.000670, -0.000883, 0.00107, -0.00107, 0.000840, -0.000712, 0.000391, -0.00103, -0.00284,\right.\\
			&&\left.\textbf{-0.00838}, -0.0740, -0.251, -0.549, -0.675, -0.405, -0.102, -0.00954, -0.000142, -0.000112, 0.0000873, -0.0000661\right\}\\
			\widetilde{\mathbf{f}}^{(7)}&=&\left\{
			-0.0000804, 0.000108, 0.000145, 0.00954, 0.102, 0.405, 0.675, 0.549, 0.251, 0.0740, 0.0168, 0.00283,\right.\\
			&&\left.\textbf{0.00103}, -0.000395, 0.000716, -0.000844, 0.00107, -0.00106, 0.000878, -0.000663, 0.000486, -0.000359, 0.000272, -0.000222\right\}\\
			\Lambda^{(8)}(\u E)&=&0.0000507 - 0.000436  \u E\\
			\mathbf{f}^{(8)}&=&\left\{
			-0.0000667, 0.0000770, -0.0000934, 0.000112, -0.000125, 0.000111, -0.0000374, -0.0000819, 0.000122, -0.000261, -0.000134, -0.00259,\right.\\
			&&\left.\textbf{-0.00705}, -0.0762, -0.280, -0.614, -0.666, -0.303, -0.0474, -0.00264, 0.000257, -0.000193, 0.000132, -0.0000907\right\}\\
			\widetilde{\mathbf{f}}^{(8)}&=&\left\{
			-0.000124, 0.000188, -0.000253, 0.00264, 0.0474, 0.303, 0.666, 0.614, 0.280, 0.0762, 0.0141, 0.00259,\right.\\
			&&\left.\textbf{0.000141}, 0.000253, -0.000112, 0.0000683, 0.0000537, -0.000125, 0.000134, -0.000118, 0.0000966, -0.0000778, 0.0000633, -0.0000541\right\}\\
		\end{eqnarray*}
	}
	Note, the eigenvectors are only defined up to an overall $(-1)$ factor.
	
\end{widetext}

\end{document}